\documentclass[aps,pra,twocolumn,epsfig,groupedaddress,showpacs]{revtex4}
\usepackage{latexsym}
\usepackage{epsfig}
\usepackage{graphicx}
\usepackage{textcomp}

\bibstyle{prsty}
\begin{document}

\title{Atom detection in a two-mode optical cavity with intermediate coupling: Autocorrelation studies}
\author{D. G. Norris}
\affiliation{ Joint Quantum Institute, Department of Physics
University of Maryland and National Institute of Standards and
Technology, College Park, MD 20742 USA.}
\author{E. J. Cahoon }
\affiliation{ Joint Quantum Institute, Department of Physics
University of Maryland and National Institute of Standards and
Technology, College Park, MD 20742 USA.}
\author{L. A. Orozco }
\affiliation{ Joint Quantum Institute, Department of Physics
University of Maryland and National Institute of Standards and
Technology, College Park, MD 20742 USA.}

\date{\today}

\begin{abstract}
We use an optical cavity in the regime of intermediate coupling between atom and cavity mode to detect single moving atoms.  Degenerate polarization modes allow excitation of the atoms in one mode and collection of spontaneous emission in the other, while keeping separate the two sources of light; we obtain a higher confidence and efficiency of detection by adding cavity-enhanced Faraday rotation.  Both methods greatly benefit from coincidence detection of photons, attaining fidelities in excess of 99 \% in less than 1 $\mu$s.  Detailed studies of the second-order intensity autocorrelation function of light from the signal mode reveal evidence of antibunched photon emissions and the dynamics of single-atom transits.
\end{abstract}

\maketitle
\section{Introduction}\label{introduction}

Atoms coupled to a finite number of modes of a cavity form a cavity quantum electrodynamic (QED) system \cite{berman94}. These objects have intrinsic interest as physical systems, and they are important for applications in quantum information science \cite{turchette95b,cirac97,gheri98,schon05}. Beyond quantum information, single atom detection with cavities is the tool of choice for the study of atom dynamics and coherent atom optics \cite{mabuchi99,munstermann99,ottl05,poldy08,khudaverdyan08}.

Cavities have also enabled the study of quantum optics effects difficult to observe in free space.  For example, Ref. \cite{hennrich05} shows a transition from antibunching to bunching in light collected from an atomic beam into a cavity mode.  The effect is visible because the cavity supports only one spatial mode out of the infinity in free space, and this ensures that the light detected from different atoms is spatially coherent.

Any measurement protocol involving individual atoms in a cavity requires information that the atom is coupled to the cavity mode.  The information manifests itself in two ways: as an increase in the amount of light in the mode owing to collection of atomic spontaneous emission, or as a decrease in the amount of light that the cavity transmits, owing to atomic absorption or dispersion.  Cavity QED enhances the rate of spontaneous emission collection, and excitation perpendicular to the cavity mode enables high rejection of background light \cite{teper06,nussmann05,fortier07}.  Detection of a decrease in driven cavity transmission requires the averaging of the intrinsic shot noise until the intensity level change is significant, and needs strong atom-cavity coupling for rapid detection  \cite{mabuchi96,aoki06,trupke07,puppe07}.  Experiments have demonstrated single atom detection using both methods for moving atoms \cite{ottl05,teper06} and trapped atoms \cite{boozer06,fortier07}.

We have developed a technique for atom detection that combines elements of driven cavity QED detection and spontaneous emission collection. We observe atoms in the mode volume of an optical cavity through the collection of spontaneous emission and Faraday-rotated forward scattering into an orthogonally polarized cavity mode that is not driven \cite{terraciano09}.  The technique allows rapid identification of highly-coupled atoms, as well as measurements of photon correlations and single-atom dynamics that are difficult to perform in free space.  This paper presents our approach, in particular detailed autocorrelation studies of the transmitted light that is basically the result of resonance fluorescence. We identify contributions from different physical processes to the autocorrelation signal in the short time (atomic decay) and long time (transit of atoms through the mode) \cite{kimble78,carmichael78,carmichael80}.

The paper is divided as follows: Section~\ref{system} introduces the system, with a simplified theoretical model in section \ref{theory} and a description of the apparatus in section \ref{apparatus}. Examples of our raw signals follow in section \ref{results}. Section \ref{analysis} explores the parameter space to establish the configuration for optimal fidelity (least chance of error in detection).  Section \ref{autocorrelations} presents a study of the autocorrelation function of the perpendicular (undriven) mode.  We conclude in section \ref{conclusions}.

\section{Detection Scheme}\label{system}
\begin{figure}
\begin{center}
\includegraphics[width=3.1in]{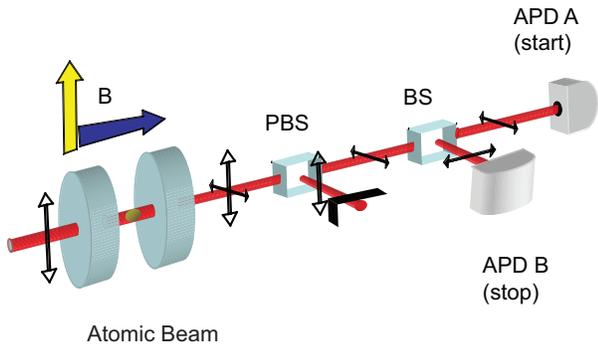}
\end{center}
\caption{(Color online). Schematic of the apparatus with
the basic optical elements necessary for coincidence measurements.
Vertically polarized laser light (parallel) drives the high finesse cavity
traversed by a continuous beam of cold $^{85}$Rb atoms from a magneto-optical trap. The output light
passes through a polarizing beam splitter (PBS) that separates the
horizontal (perpendicular) polarization, sending it to a nonpolarizing beam
splitter (BS) for coincidence measurements using two avalanche
photodiodes (APD). The direction of the magnetic field (B) inside the cavity can be parallel to V ($\pi$-polarized drive) or parallel to the direction of propagation ($\sigma$-polarized drive).
\label{figure2}}
\end{figure}

The basic components of our system are typical of cavity QED setups (see Fig.~\ref{figure2}).  We couple a linearly polarized laser (the drive) to a TEM$_{00}$ mode of a Fabry-Perot optical cavity resonant with the $D_2$ line in $^{85}$Rb.  Intersecting the cavity axis at near-normal incidence is a slow-moving beam of neutral $^{85}$Rb atoms, each of which interacts with the cavity mode for some time before leaving.  Because of the large difference in time scales (5 $\mu$s transit time versus 26 ns excited state lifetime), the atoms become excited and scatter the drive light many times while in the mode volume.

The crux of our detection scheme is the separation of the cavity output field into two distinct components relative to the polarization of the exciting laser: one parallel (driven mode) which is little affected by the transiting atoms, and the other perpendicular (undriven mode) which in the ideal case is populated only by scattered light from the atoms (see Ref. \cite{birnbaum05} for the pioneering use of this configuration).  The two polarization modes are degenerate (simultaneously resonant in the cavity), so we can use the same cavity for both excitation and signal collection.

\begin{figure}
\begin{center}
\includegraphics[width=3.1in]{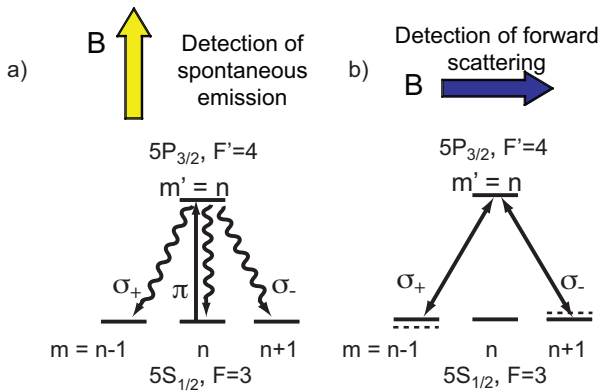}
\end{center}
\caption { (Color online). a) A linearly polarized driving field excites the atom, but a transition to a ground state level with different $m$ results in emission in the orthogonal polarization. b) Simplified diagram of $^{85}$Rb levels relevant to Faraday rotation in an applied magnetic field.
\label{figure1}}
\end{figure}

We consider two different configurations for excitation. The first has a weak magnetic field parallel to the incident polarization.  The light drives $\pi$ ($\Delta m$=0) transitions in the atoms in this basis (see Fig.~\ref{figure1}a).  An excited atom can return to its original ground state $m$-level through spontaneous or stimulated emission into the parallel (driven) mode; however, there is also some probability to relax into a different $m$-level through a $\sigma^{+}$ or $\sigma^{-}$  ($\Delta m=\pm 1$) spontaneous emission transition, which produces light that is circularly polarized with respect to the magnetic field axis and may be collected by the perpendicular (undriven) cavity mode.
Although the cavity coupling strength is modest, the amount of collected light increases measurably from that in free space \cite{terraciano07a}.

The second configuration has a weak magnetic field along the cavity axis, while keeping the same polarization arrangement (see Fig.~\ref{figure1}b).  The linearly polarized drive can then be considered as the sum of two circular components with different indices of refraction due to the Zeeman shift. This results in a Faraday rotation of the drive into the perpendicular mode.  Though ordinarily the Faraday rotation imparted by a single atom in a weak magnetic field is vanishingly small, the cavity finesse allows the light to compound multiple rotations before escaping, and the effect becomes appreciable.

A polarizing beam splitter (PBS) sends the two modes to separate detectors, which can resolve individual photon events.  From the record of photon detections, we analyze the statistics of emission to look for signs of single atom transits.  One advantage of collection into a cavity mode is that we can spatially filter most sources of background light.  However, there are remaining contributions to the perpendicular mode from cavity mirror birefringence and background light that cannot be completely eliminated. The optimal parameters for operation minimize the background influence on the atom detection confidence.

\section{Theory}\label{theory}

The coupling of a single photon to a single atom in optical cavity QED takes place through electric dipole transitions at the Rabi frequency $g_{0}\equiv\vec{\mu}\cdot\vec{E}/\hbar$, where $\vec{\mu}$ is the atomic dipole moment and $\vec{E}=\sqrt{\hbar \omega/2 \epsilon_{0}V}\hat{x} $ is the electric field amplitude of a single photon in the cavity mode volume $V$and polarization $\hat{x}$. The coupling constant $g$  varies in space with the cavity mode function.  For a TEM$_{00}$ Gaussian standing-wave mode of wavelength $\lambda$ in a cavity with axis along the $\hat{z}$ direction, the coupling takes the form:
\begin{equation}
g(x,y,z)=g_0\cos(kz)e^{-(x^2+y^2)/w_0^2},
\label{g}
\end{equation}
where the amplitude $g_0$ depends on the transition dipole moment and Clebsch-Gordan coefficients, $k=2\pi/\lambda$ and $w_0$ is the mode waist.

\begin{figure}
\begin{center}
\includegraphics[width=3.1in]{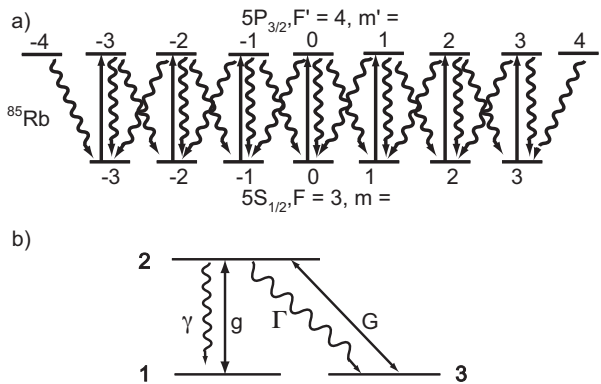}
\end{center}
\caption{a) Diagram of $^{85}$Rb levels relevant to the detection. b) Energy level diagram of the simplified model.
\label{figure1extra}}
\end{figure}

The real structure of the atomic hyperfine levels in our experiment is complex (Fig.
\ref{figure1extra}a) with a wide range of values for $g_0$. We present here a three-level model (Fig. \ref{figure1extra}b) following Ref.~\cite{terraciano07a}: two ground states ($\left | 1 \right >$ and $\left | 3 \right >$)  and one excited state ($\left | 2 \right > $). We call the coupling constants $g$ and $G$ for the two modes and the decay rates $\gamma$ and $\Gamma$ for the two channels.  The total decay rate of the population inversion is the sum,

\begin{equation}
\gamma_{tot}=\gamma+\Gamma.
\end{equation}
The rates are related by $g/G=\gamma/\Gamma$, where the ratio depends on the Clebsch-Gordan coefficients for the specific transitions involved.
The number of photons needed to saturate the driven atomic transition is:\begin{equation}
n_{0}=\frac{\gamma_{tot}^2}{3g_0^2}
\end{equation}
which includes a geometric factor from the Gaussian standing wave \cite{drummond81}.

Together with the saturation photon number $n_0$, we take as a figure of merit the single-atom cooperativity $C_1$, a dimensionless parameter which characterizes the influence of a single atom in the driven mode with decay rate $\kappa$: \begin{equation}
C_1 = \frac{g_0^2}{\kappa\gamma_{tot}}.
\label{c1}
\end{equation}
An equivalent parameter for the undriven mode is:
\begin{equation}
\widetilde{C}_1=\frac{G_0^2}{\kappa\gamma_{tot}}.
\end{equation}
We define a total cooperativity parameter $C$ for $N$ atoms in the two-mode system, which is slightly different from that for two-level atoms ({\it i.e.}  $C_{1}N$) \cite{lugiato84}:
\begin{equation}
C=\frac{C_1 N}{1+2\widetilde{C}_1}.
\end{equation}

We consider a cavity resonant with both atoms and input drive, and $N$ atoms stationary and maximally-coupled to the mode.  For a driving intensity of $Y= \left <n \right>/n_0$ (representing the mean steady-state number of photons inside the empty cavity, normalized to the saturation number) and the three-level atom, two-mode cavity model discussed above, the steady-state photon number with atoms in the driven ($X_{\|}$) and perpendicular ($X_{\bot}$) modes is \cite{terraciano07a}:
\begin{eqnarray}
X_{\|}&=&Y\frac{1}{(1+2C)^2}\approx Y\left(1-4C\right), \label{xpar}\\
X_{\bot}&=&Y\left[\frac{2\widetilde{C}_1}{1+2\widetilde{C}_1}\right] \left[\frac{C}{\left(1+2C\right)^2}\right] \approx Y 2 \widetilde{C}_1C, \label{xperp}
\end{eqnarray}
where the lowest order approximations are valid for $2C\ll1$ and $2\widetilde{C}_{1}\ll1$.  The signal $X_{\bot}$ increases linearly with the number of atoms and contains no additional terms, indicating that any amount of light is the result of atomic emission.  This is in contrast to the parallel mode signal $X_{\|}$ which is the difference of the drive, determining the shot noise, and a small number proportional to $2C$ that carries the information of the atoms.

To include the effects of Faraday rotation, we consider an atom with one excited state ($m'=0$) and three ground states ($m=\pm 1, 0$). (See Fig~\ref{figure1}b). A magnetic field $B$ along the direction of light propagation lifts the ground state degeneracy through the Zeeman effect, resulting in a shift in the dispersion curves and a difference in the phase velocities of the two circular components of light, causing Faraday rotation \cite{budker02}. We excite the system with vertically polarized light, which is an equal superposition of right- and left-circularly polarized light. The light interacts with the atom and becomes elliptically polarized; the plane of polarization rotates through an angle. The light then passes again through the atom after reflection on the mirror and compounds the rotation. When the light exits the cavity, the total rotated angle is that of a single pass multiplied by the number of times the light passes through the atom, indicated here by the presence of the cooperativity $C$:
\begin{equation}
\phi=\frac{2g_{L} \mu_{B} B/\hbar \gamma_{tot}}{1+(2g_{L} \mu_{B} B/\hbar \gamma_{tot})^2}C,
\label{phi}
\end{equation}
where $g_{L}$ is the Land{\'{e}} factor and $\mu_{B}$ is the Bohr magneton.

The addition of this effect changes Eq. \ref{xperp} into:

\begin{equation}
X_{\bot}= Y \left( 2 \widetilde{C}_1C + |\phi |^2  \right). \label{xperp2}
\end{equation}
Equation \ref{xperp2} shows that the signal in the perpendicular mode now consists of two parts: a background level of spontaneous emission (which may be considered as $(\sigma^{+}-\sigma^{-})$ polarization if the drive is $(\sigma^{+}+\sigma^{-})$, in full analogy to the basis with $\pi$ drive), and Faraday-rotated drive which increases from zero to a maximum value of $X_{\bot {\rm Faraday}}=YC^2/4$ when $B=\hbar \gamma_{tot}/(2g_{L} \mu_{B})$.  For simplicity in what follows we treat the atom-cavity coupling in the basis of $\pi$ drive, which must agree with the total coupling in the basis of $\sigma$ drive when the ground-state populations are appropriately transformed (though the values of $g_0$ and $C_1$ for the individual transitions are generally different).

We analyze the two-time intensity autocorrelation function of the light from the perpendicular mode to understand the dynamics of photon emissions that contribute to the steady-state value of $X_{\bot}$.  We do not distinguish between the two components in Eq. \ref{xperp2}, treating both as arising from near-resonant scattering of incident light (resonance fluorescence) in which the cavity serves only as a mode in which to collect the light.

For a stream of photons scattered from a single two-level atom undergoing stationary, resonant excitation, the intensity autocorrelation function takes the familiar form for resonance fluorescence antibunching:
\begin{equation}
g^{(2)}_{A}(\tau) = 1-e^{-(3\gamma/4)\tau}(\cosh \delta\tau + \frac{3\gamma}{4\delta}\sinh \delta\tau),
\label{antib}
\end{equation}
where $\delta=(\gamma/4)\sqrt{1-8Y}$ is a function of the drive intensity $Y$ \cite{carmichael93book}.  The antibunching referred to here means an increasing slope for increasing times: $dg^{(2)}(\tau)/d\tau|_{\tau>0}>0$.
The condition $g^{(2)}(0)<1$ indicates that the photons obey sub-Poisson statistics, and is not a necessary condition for the photons to exhibit antibunching.

The modification of Eq. \ref{antib} for the case of an atomic beam passing through a localized excitation region appears in Ref. \cite{kimble78}, which adds a number of effects: i., the multiplication of Eq. \ref{antib} by a window function to include the shape of the mode function encountered by each transiting atom;  ii., the scaling of Eq. \ref{antib} by the inverse of the steady-state mean number of interacting atoms, $\overline{N}$; iii., the addition of a term $|g^{(1)}_A(\tau)|^2$ to include the beating of fields emitted by different atoms; iv., the addition of a constant offset of one arising from uncorrelated emissions from different atoms; and v., the reduction of the correlation factor due to the presence of completely uncorrelated background light.  Further generalizations appear in Ref. \cite{carmichael78,carmichael80}, where the authors also consider the product of the fields that come from atomic emission with a temporally uncorrelated but spatially mode-matched field from background light.

Our system is more complicated than either of these cases because the birefringence background, as a component of the driving laser, is temporally coherent and completely mode-matched with the atomic emission.  This gives a non-zero expectation value for the product of atomic and background fields, and adds a number of complicated terms to the correlation function which become increasingly important with increasing drive intensity.  Such terms should decay to zero on the time-scale of the atomic lifetime, however, and are unimportant for weak driving, so we omit them from the present analysis and concern ourselves with understanding the atomic motion and number fluctuations, which manifest themselves mainly in the long tails of the correlation function.

We are able to describe the total correlation function as:
\begin{equation}
g^{(2)}(\tau) = 1 + \frac{1}{(1+R_b/R_s)^2}\frac{f(\tau)g^{(2)}_{A}(\tau)}{\overline{N}}+F(\tau;R_b/R_s,\overline{N})
\label{g2}
\end{equation}
where $R_b$ and $R_s$ are the average count rates for background and signal, respectively, and
\begin{equation}
f(\tau) = [\cos(\Omega\tau)e^{-\beta\tau}+1]e^{-(\tau/T)^2}
\label{foftau}
\end{equation}
is an empirically-determined window function describing motion of the atom through the cavity mode with Gaussian temporal width $T$, where a small tilt of the atomic beam with respect to the cavity axis normal results in motion through the standing-wave lobes at frequency $\Omega$.  These oscillations decay at a rate $\beta$ much faster than the transit time, due to the spread in velocities from the atomic source.  The function $F$ corresponds to any additional terms related to the beating of signal from different atoms or signal with background, which at low intensities is dominated by partially mode-matched scattered light from outside the cavity.  Since this term represents classical noise sources, it can only contribute light which is bunched around $\tau=0$.  Any observation of antibunching in our signal must come from the single-atom term $g^{(2)}_{A}(\tau)$.  An explicit observation of $F$ would require $\overline{N}\gg1$ since its width is comparable to that of the antibunching signal  \cite{puppe07}.

\section{Apparatus}\label{apparatus}

The apparatus consists of two main
components: the source of atoms and the cavity.
A titanium sapphire laser (Ti:Sapph) provides most of the light
needed for the experiment at 780 nm. The laser linewidth and
long-term lock are controlled using a Pound-Drever-Hall (PDH)
technique on saturation spectroscopy of $^{85}$Rb.  The
excitation beam of the cavity passes through an electro-optical
modulator to imprint frequency sidebands separated by
150 MHz from the carrier. One sideband is near resonance with the atomic transition and the cavity, becoming the system drive, while the carrier and the other sideband are far detuned and reflect from the front cavity mirror. Changing the sideband frequency allows us to probe the frequency response of the resonant cavity-atom system without misalignment of the input laser coupling.

Polarization elements (better than $5 \times 10^{-5}$) and mode-matching optics prepare the driving laser before it enters the cavity. A lens at the exit collimates the beam and a half-wave plate (HWP) aligns the polarization to a calcite PBS which separates the parallel (driven) and perpendicular (undriven) modes. The perpendicular mode passes through a second beam splitter which can divide the light between two avalanche photodiodes (APD) (Perkin Elmer SPCM-AQR-12 and -13) for single or coincidence measurements. A series of filters, telescopes and apertures before the detectors remove background light. The APD electronic output pulses then go to a time-stamp unit in a computer or to a correlator \cite{foster98}. The system has a  23\% measured photon detection efficiency for the APDs and optical paths.

The Fabry-Perot cavity has a mirror separation of 2.2 mm and a 1/$e$ field TEM$_{00}$ mode waist of 56 $\mu$m. The input mirror transmission (15~ppm) is lower than the output  mirror transmission (300~ppm) by a factor of 20 to
ensure that most of the signal escapes from the cavity on the
detector side. The decay rate for the cavity is $\kappa/2\pi=3.2  \times 10^6$ s$^{-1}$ with finesse of 11,000.  The cavity length is kept resonant with the $F=3 \rightarrow F'=4$
transition of the $D_2$ line of $^{85}$Rb at 780 nm.  The birefringence splitting of the two polarization modes is less than 500 kHz. The cavity length is stabilized by the PDH method with light
derived from an auxiliary laser at 820 nm.  A grating and interference filters separate the 780 nm signal light and 820 nm locking light on the cavity output.

A rubidium dispenser delivers Rb vapor to a magneto-optical trap
(MOT) in a stainless steel chamber above the vacuum chamber that houses the
cavity. Both are kept under vacuum with pressures below $10^{-8}$ Torr.  We use a six-beam configuration with 1/$e$
power diameter of 20 mm and 10 mW per beam.  A second laser
repumps the atoms that fall out of the cycling transition in the
trap. Combinations of acousto-optical modulators (AOM) permit
independent settings of the frequency and amplitude of all laser
beams, controlled by computer. A pair of
coils generates a magnetic field gradient of 10 G/cm,
and three sets of independent coils zero the magnetic field at the
trapping region. Three extra coils around the cavity region allow for configuring a constant field perpendicular or parallel to the axis of the cavity without disturbing the location of the MOT.

The retro-optics of the vertical arm of the MOT  (inside the vacuum chamber)  have
1.5 mm-diameter holes on axis. This creates an imbalance in the light pressure of the vertical cooling beam such that cold atoms are pushed down into the cavity, producing a low velocity atomic beam (LVIS) \cite{lu96}.   The atomic beam intersects the cavity mode about 8 cm below the trapping region.  The mean speed of the atoms passing through the cavity mode is approximately 15 m/s.  Ref. \cite{lu96} shows that the distribution of longitudinal speeds is considerably wider than that of a sample at the Doppler temperature, while the spread in transverse velocities results primarily from the geometric collimation.

\begin{table}[htdp]
\caption{Clebsch-Gordan (CG) coefficients, single-atom coupling constants $g_0$, single-atom cooperativities $C_1$, and saturation photon numbers $n_0$ for different transitions in the $D_{2}$ line of $^{85}$Rb from $F=3$ to $F'=4$, for different drive polarizations($\pi$ or $\sigma^{+}$).}
\begin{center}
\begin{tabular}{|c|c|c|c|c|c|c|}
  \hline
  &$F=3$&$F'=4$ &CG&
 $g_0/2\pi$ [MHz] & $C_1$ & $n_0$ \\
  \hline
  $\pi$&$m=0$&$m'=0$&$-\sqrt{2/7}$ & 1.5 & 0.12 & 5.3 \\
  &$m=3$&$m'=3 $&$-\sqrt{1/8}$& 0.99 & 0.053 & 12 \\
  \hline\hline
    &$m=0$&$m'=1$ & $\sqrt{5/28}$&1.2 & 0.075 & 8.5 \\
    $\sigma^{+}$&$m=3$&$m'=4$ &$ \sqrt{1/2}$&2.0 & 0.21 & 3.0 \\
    &$m=-3$&$m'=-2$ & $\sqrt{1/56}$ & 0.38 & 0.0075 & 85\\
    \hline
\end{tabular}
\end{center}
\label{default}
\end{table}

The single-atom dipole coupling frequency for the driven cavity
mode depends on the Clebsch-Gordan coefficient for the particular states involved.  Table 1 gives the expected range of values for $g_0$, $C_1$, and $n_0$, for transitions between the $F=3$ ground state and the  $F'=4$ excited state.  The strongest coupling for $\pi$-polarized light occurs for $m=0$ and the weakest for $m=3$.  The atoms enter the cavity in a distribution of ground states that is not centered around $m=0$.  The total atomic decay rate is $\gamma_{tot}/2\pi=6 \times 10^6$ s$^{-1}$.  We write all intensities normalized to the $m=0$, $\pi$-polarized saturation intensity, $n_0=5.3$ photons.

We find the peak count rate for the perpendicular mode with Faraday rotation by scanning the laser and cavity together across the atomic resonance.  We take all measurements at the location of the peak ($|B|$=3.3 Gauss), which is shifted by about 3 MHz from the $F=3, m=0 \rightarrow F'=4, m'=0$ transition.

For evaluating photon statistics, we detect the perpendicular mode light on a single APD for 300 s at different drive intensities.  The detection records are stored on computer for post-processing and analysis, with a resolution of 4 ps.  For measuring the intensity autocorrelation, we split the light between two APDs to eliminate distortions from detector dead time and after-pulsing.  These are also recorded for 300 s at each intensity, at the same atomic beam density used for the single APD measurements.  The autocorrelation is formed by making a histogram of the time between coincidences on the two detectors, out to a delay of $\pm$10 $\mu$s with a bin resolution of 10 ns.  We take as normalization the mean number of counts per bin, based on the long-time average count rates.  Measurements for calibrating Faraday rotation were taken separately, at a similar atomic beam density.

\section{Preliminary data analysis}\label{results}
\subsection{Faraday rotation}

The Stokes parameter formalism \cite{labeyrie01}
 allows a measurement of the rotation angle $\phi$ if the amplitudes of the electric field in the two polarizations $\varepsilon_{\perp}$ and $\varepsilon_{\parallel}$ are known:

\begin{equation}
|\phi|= \frac{|\varepsilon_ {\parallel} \varepsilon_{\perp}|}{|\varepsilon_{\parallel}|^2 - |\varepsilon_{\perp}|^2} \approx \left | \frac{\varepsilon_{\perp}}{\varepsilon_{\parallel}} \right |,
\end{equation}
where the approximation is made for $|\varepsilon_ {\parallel}|\gg |\varepsilon_{\perp}|$.
We measure the count rates in the perpendicular mode as a function of applied magnetic field along the cavity axis to determine the location of maximum Faraday rotation. The peak rotation is different depending on the sign of the magnetic field, as the atoms experience different optical pumping by the pushing beam from the MOT, and we do not otherwise optically pump before the cavity.  We obtain a maximum of $|\phi|=0.035 \pm 0.007$ rad from the measured increase in the count rate with a field magnitude of 3.3 Gauss.  To make explicit the effects of the cavity, we estimate the effective number of maximally-coupled atoms by using Eq. \ref{xpar} and measuring the driven mode transmission.  This was greater than 0.95, giving a bound of $C<$0.01 or $N<$0.1 effective maximally-coupled atoms (for $\pi$ transitions from $m=0$).  Calculating the free-space absorption length for 0.1 atoms and our cavity waist, we would expect a maximum rotation of the order of 3 $\mu$rad \cite{budker02}.  The enhancement by a factor of approximately 10,000 is a result of the cavity finesse.

\subsection{Photons per atom}

The formula of Mandel relates the photon number distribution $P(n)$ with the atom number distribution $P_{atom}(m)$:

\begin{equation}
P(n)=\sum_{m}P(n|m)P_{atom}(m),
\label{mandelformula}
\end{equation}
where $P(n|m)= (\alpha m)^{n}\exp{(-\alpha m)}/n!$ is the conditional probability of detecting $n$ photons when there are $m$ atoms in the cavity volume, each contributing a mean of $\alpha$ photons to the signal with a Poisson distribution of number \cite{wilzbach08}. This shows that the super-Poisson fluctuations of the light arise from the combined Poisson fluctuations in atom number and photons scattered, with $\alpha$ an extractable parameter.

We follow the method of Ref.~\cite{teper06} to relate the mean and mean squared photon number ($\langle n \rangle $, $\langle n^{2}\rangle $) to $\alpha$:
\begin{equation}
\frac{\langle n^2 \rangle}{\langle n \rangle}-1=\langle n \rangle g_{aa}+\alpha,
\label{line}
\end{equation}
where the atom-atom correlation function is $g_{aa}=\left ( \langle m^2 \rangle - \langle m \rangle \right )/ \langle m \rangle ^2$, which for Poisson atomic fluctuations reduces to $g_{aa}=1$.

\begin{figure}
\begin{center}
\includegraphics[width=3.1in]{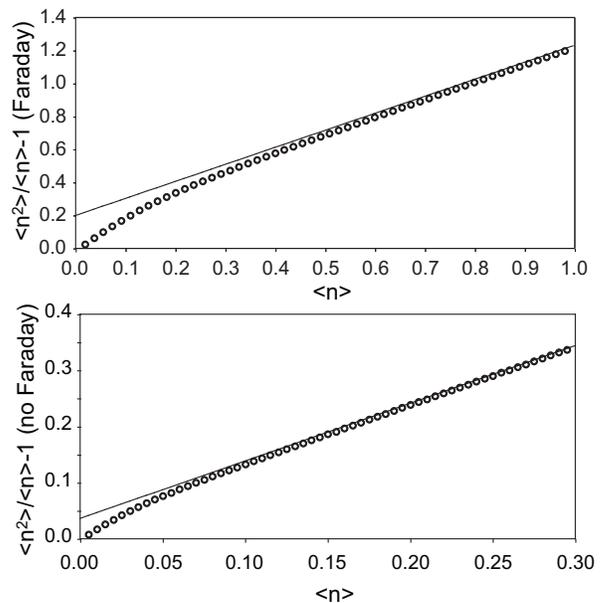}
\end{center}
\caption{Examples of data (circles) and least-squares fits (continuous line) to Eq. (\ref{line})  for Faraday rotation (top)  and for no Faraday rotation (bottom) as a function of average number of photons in a time bin. }
\label{figure4}
\end{figure}

Figure \ref{figure4} shows a detail of the plot of the experimental values for the left side of equation (\ref{line}) with the two different detection methods: Faraday rotation on the top and no Faraday rotation on the bottom. We calculate the statistics based on time series of 300 s that we bin in time.  Longer bin sizes have a larger average of counts (horizontal axis). The  intercept  gives the value of $\alpha$.  A linear fit for bins from 50 $\mu$s to 100 $\mu$s ($\langle n \rangle=$ 1.81 to 3.63, off the scale of the plot in Fig.\ref{figure4}) in the Faraday case gives a vertical axis intercept of $\alpha=0.196 \pm 0.003$ photons detected per atom, and a slope of 1.033 $\pm$ 0.001. The intercept in the no Faraday case (for a fit from $\langle n \rangle =$ 0.50 to 1.00) is $\alpha=0.036 \pm 0.002$ photons detected per atom and a slope of $1.026 \pm$ 0.003. The drive in both cases is 0.4 $n_{0}$.  The value of the slopes indicates that the assumption that the atoms follow Poisson statistics is well justified.

\begin{figure}
\begin{center}
\includegraphics[width=3.1in]{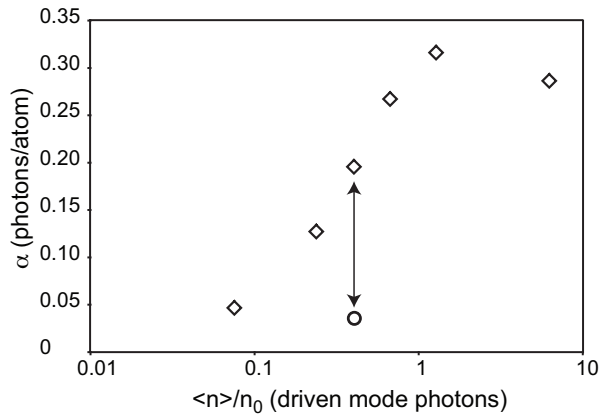}
\end{center}
\caption{Counts detected from a single atom (rhombs for Faraday, circle for non Faraday) as a function of intensity of the exciting laser normalized by the saturation photon number. The arrows indicate the points extracted from the plots on Fig.~\ref{figure4}. The error bars are smaller than the size of the data points. }
\label{figure5}
\end{figure}

Figure~\ref{figure5} shows the extracted values of $\alpha$ as a function of cavity photon number (driving intensity) spanning more than two orders of magnitude. There are data points (rhombs) that come from Faraday rotation and one with spontaneous emission alone (circles). The latter is less efficient by almost a factor of five.  $\alpha$ increases linearly with drive until atomic saturation intensity $(n\approx n_{0})$, where also optical pumping effects start to enter. Each point on Fig.~\ref{figure5} comes from a least-squares fit similar to those in Fig.~\ref{figure4}. The count rates in the perpendicular mode and the values of  $\alpha$ give an atomic flux of approximately 160,000 atoms $s^{-1}$.

\section{Analysis and optimization of detection}\label{analysis}

Three identifiable sources of background counts contribute to the measurement of the perpendicular mode: intrinsic detector dark counts (approximately 300 $s^{-1}$ each), scattered light from the MOT beams (approximately 2000 $s^{-1}$), and the birefringence of the cavity (less than $5 \times10^{-5}$). The light from the MOT dominates the background count rate at low driving intensity; however, at higher intensity, light from cavity birefringence is the main contributor to the degradation of the signal-to-background ratio.
A single photon escaping in the perpendicular mode most likely comes from an atom in the cavity; however,  there is at best still a 4\% probability that it comes from  background
counts.  We suppress the probability of a false atom detection by requiring
photon coincidences in a time window smaller than
the transit time of an atom across the cavity mode \cite{aoki06}.  It is possible to implement this in real time using an electronic coincidence counter.  The first perpendicularly polarized photon detection (``start'') opens a gate of variable width which allows counting pulses from the ``stop'' APD.

\begin{figure}
\begin{center}
\includegraphics[width=3.1in]{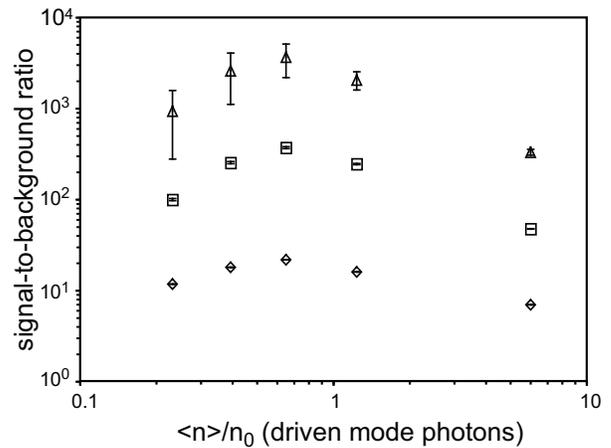}
\end{center}
\caption{Signal-to-background ratio as a function of normalized driving laser intensity for single detection (open rhombs), double coincidence detection (open squares), and triple coincidence detection (open triangles) in a 1 $\mu$s window with Faraday rotation of the drive.}
\label{figure7}
\end{figure}

Figure~\ref{figure7} shows the signal-to-background ratio calculated from time series ({\it{a posteriori}} in this case) for the rate of single photon detection (open rhombs),  two-photon coincidence (open squares), and three-or-more-photon coincidence events (open triangles) in a 1 $\mu$s window as a function of driving intensity. The maximum occurs at the same driving intensity for all three. Two photons within a 1 $\mu$s window  improves the signal-to-background ratio by more than an order of magnitude compared to detection of single photons.  The  three-photon coincidence gives a better ratio but significantly decreases the rate of detection as exemplified by the error bars associated with each point. The three sets achieve their maxima just before the atomic saturation intensity, due to background counts from cavity birefringence that continue to increase linearly with drive. The contribution to the coincidences from detector afterpulsing is small  (less than $1\%$) and does not affect the results.

To evaluate the probability of making a false detection for various coincidence gate times, we extract the fidelity of detection $F$ versus gate length for two-photon coincidence based on the time record. We calculate the waiting time distribution of photon arrivals on a single APD by making a histogram of the time separation between consecutive detections. The integral of the waiting time distribution gives the number of two-photon coincidences $c$ in a given time window.  We calculate $c$ with and without atomic flux to obtain:
\begin{equation}
F = 1 - c_{without}/c_{with}.
\label{fidelity}
\end{equation}

The fidelity (one minus the probability of error) reaches more than 99\% at 0.1 $\mu$s, and 99.7\% at 1 $\mu$s with Faraday rotation, but only 96.7\% at 1 $\mu$s with detection of spontaneous emission alone.  The fidelity decreases at longer times.  The fidelity is optimal for times between 1 and 5 $\mu$s, determined by the distribution of atom transit times.

\section{Intensity autocorrelation}\label{autocorrelations}

We ensure that our coincidence detection scheme is sensitive to single atoms by measuring the intensity autocorrelation function ($g^{(2)}(\tau)$) of the perpendicular mode under very weak driving intensity at the same atomic beam density used during detection measurements.  If we consider the light in the perpendicular mode as coming from resonance fluorescence, the antibunching can only arise from photon pairs from the same atom.  The observation of antibunching in our signal is strong indication of our ability to detect photon coincidences from a single atom in less than 1 $\mu$s.

\begin{figure}
\begin{center}
\includegraphics[width=3.1in]{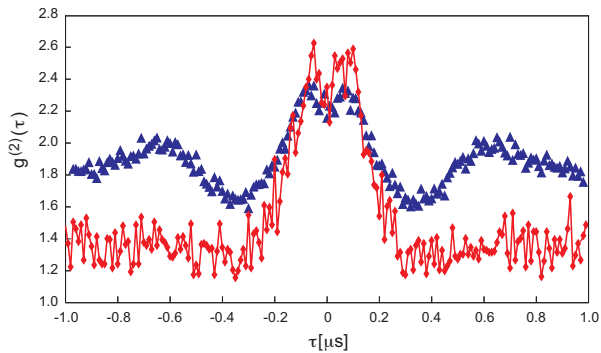}
\end{center}
\caption{(Color online) Intensity autocorrelation function ($g^{(2)}(\tau)$) of the perpendicular mode with (triangles) and without (diamonds) Faraday rotation, at a driving intensity of 0.4 $n_0$.}
\label{figure12}
\end{figure}

Figure~\ref{figure12} shows two examples of an autocorrelation function around $\tau=0$ with and without Faraday rotation for the same driving intensity (0.4 $n_0$). The antibunching is visible and lasts for a time of the order of the excited state lifetime (26 ns). The area under the curve is larger with Faraday rotation, indicating a substantial increase in photon flux from individual atoms.  The antibunching is less pronounced without the rotation because of the lower signal-to-noise ratio.

\begin{figure}
\begin{center}
\includegraphics[width=3.1in]{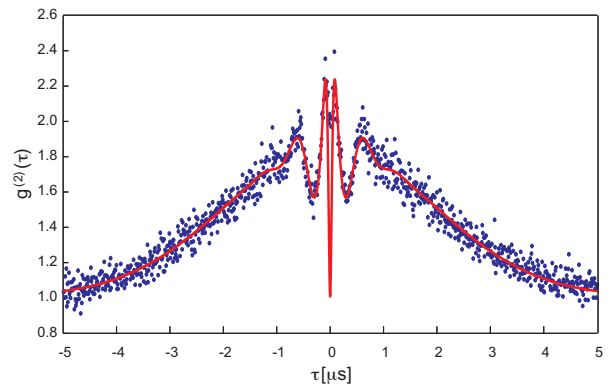}
\end{center}
\caption{(Color online) Intensity autocorrelation function ($g^{(2)}(\tau)$) of the perpendicular mode at 0.24 $n_{0}$. The continuous line is the theoretical fit to Eq.~\ref{g2}.}
\label{figure11}
\end{figure}

Since we record the photon detections as continuous time series, we can also look at the long-term behavior of the autocorrelation in the context of resonance fluorescence as presented in Ref.~\cite{carmichael78,carmichael80,kimble78}. We fit the data to Eq. \ref{g2} with free parameters $\overline{N}$, $T$, $\beta$, and $\Omega$, excluding the term $F(\tau)$ and the region $|\tau|<$50 ns where it is important.  We find excellent agreement in this outer region (reduced $\chi^{2}$=1.03 for points out to $\pm$5 $\mu$s).  Fig. \ref{figure11} shows the fit to the data taken at a driving intensity of 0.24 $n_{sat}$.  The damped oscillation at $\Omega$=1.5 MHz corresponds to motion through the standing wave with a period of 0.67 $\mu$s (a velocity of 0.58 m/s along the cavity axis) and a damping time of $1/\beta$=0.29 $\mu$s.  Monte Carlo simulations of the transit show good agreement with this damping time based on the geometric collimation of transverse velocities from the source \cite{lu96}.  The Gaussian background gives a mean atom number of $\overline{N}$=0.88, and a 1/$e$ waist of $T$=2.7 $\mu$s (a mean velocity of 14.7 m/s across the Gaussian mode).  The ratio of the velocities gives a beam tilt of 2.3$^\circ$ with respect to the cavity axis normal. Although we take care to align the cavity under the exit hole of the atomic source, the cavity mode is not necessarily centered on the two mirrors, and the transverse beam width allows for small inclinations of the beam.

\begin{figure}
\begin{center}
\includegraphics[width=3.1in]{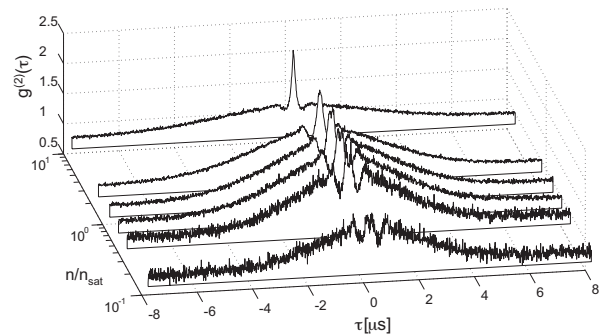}
\end{center}
\caption{Intensity autocorrelation function ($g^{(2)}(\tau)$) of the perpendicular mode for different values of photon number (intensity) in the driven mode.}
\label{figure10}
\end{figure}

The full structure of the correlation function depends not only on the above parameters (properties of the atomic beam), but also on the intensity of the drive, which can change the atomic response or increase the amount of background light from cavity birefringence. Fig. \ref{figure10} shows the evolution of the autocorrelation function as the driving intensity increases by two orders of magnitude.  We observe a transition from antibunching to bunching with higher drive.  The behavior comes from the term $F(\tau)$ in Eq. \ref{g2} which includes beating of the birefringence light with the atomic emissions.  This is in contrast to the transition seen in Ref. \cite{hennrich05}, where the atomic density is increased such that the constant $|g^{(1)}_A(\tau)|^2$ term (part of our $F(\tau)$) becomes visible. On separate experiments at low intensity we have also followed the disappearance of the antibunching as we increase the number of atoms.

\begin{figure}
\begin{center}
\includegraphics[width=3.1in]{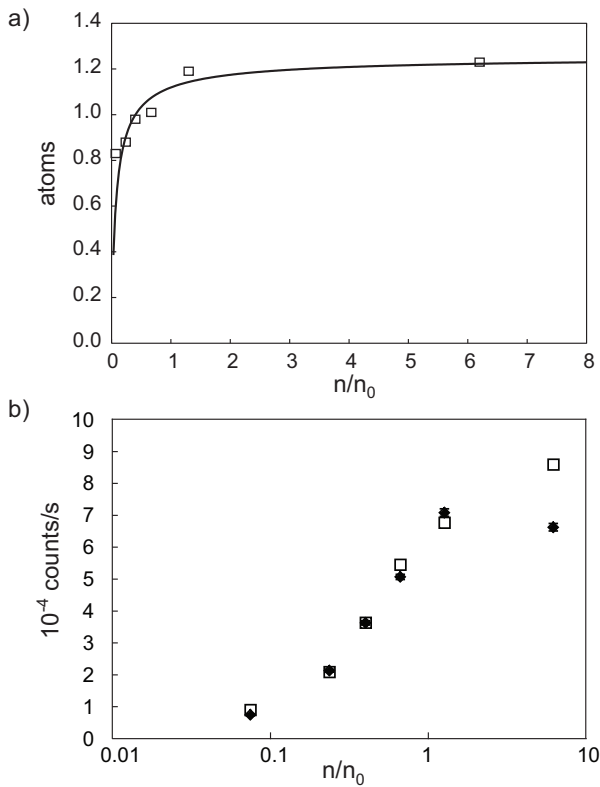}
\end{center}
\caption{a) Measured mean number of interacting atoms, and theory curve showing the expected saturation.  b) Measured (filled rhombs) count rates in the perpendicular mode together with those predicted from the extracted mean number of atoms (open squares), photons per atom, and transit time.}
\label{figure13}
\end{figure}

Although the actual density of atoms is constant, the effective excitation volume increases with power, as atoms near the nodes and wings of the mode start to interact \cite{drummond81}.  This process reaches a maximum above saturation intensity, when the only unsaturated atoms are too weakly coupled to emit into the mode.
Fig. \ref{figure13}a. shows the extracted values of $\overline{N}$ versus driving intensity for the autocorrelations in Fig. \ref{figure10}.  The line shows the expected shape of the saturation, assuming that the probability of collecting a photon from an atom is proportional to the product of a constant $g^2(x,y,z)$ for collection and a saturating strength of excitation ($[(n/n_0)g^2(x,y,z)]/[1+(n/n_0)g^2(x,y,z)]$), which is consistent with the expression $Y2\widetilde{C}_1C$ in Eq. \ref{xperp}, modified for strong driving.  We obtain the theoretical line by integrating this product over a volume much larger than the cavity mode, and using a least-squares fit to scale the vertical axis and $n_0$.

Using the extracted parameters of mean atom number and mean transit time from the two-APD correlation measurements, together with the extracted values of $\alpha$ from the one-APD measurements, we can predict the expected macroscopic count rates as $R_s=\overline{N}\alpha/2T$.  Fig. \ref{figure13}b. shows that we obtain excellent agreement for low intensities, while the highest intensity drive (largest birefringence background) disagrees significantly.  This comes from the large contribution from the beating between background and signal, which generates the large central peak in the measured autocorrelation.

We emphasize that although the mean number of interacting atoms is approximately one, this is not the same as the effective number of maximally-coupled atoms, since most are weakly coupled.  To explore this numerically, we use the work of Carmichael and Sanders \cite{carmichael99}.  We distribute atoms randomly and uniformly across the cavity mode function such that the mean density is 0.88 atoms within two mode waists (similar to the measured value for low driving intensity).  Allowing the atom number to fluctuate with a Poisson distribution, we sum the individual coupling strengths for each configuration of atoms and weight by the atomic number distribution to obtain the mean steady-state effective number of atoms, which in this case is 0.04.  Using the mean value of $C_1$=0.09 for $\pi$ transitions from the ground state, this gives a driven mode transmission of approximately 0.99 from Eq. \ref{xpar}, consistent with the experimental conditions for single atom detection. A rapid decrease in probability above one effective atom indicates that we operate in a density regime of single-atom coupling, with a low probability of having two atoms simultaneously coupled by more than half of the maximum value.   This numerical result together with the antibunching in the perpendicular mode counts confirms that the measured signals indeed arise primarily from single-atom emission bursts.

\section{Conclusion}\label{conclusions}
A two-mode cavity QED system is an excellent single-atom detector. The detection of a single photon out of the undriven mode signals the arrival of an atom in the cavity, and the detection of a second photon within 1 $\mu$s
confirms the presence of an atom with more than 99\% confidence. The method can operate over
a wide range of photon fluxes and works for atoms traveling as fast as 15 m/s across the mode.  Because the cavity is frequency-selective and has a short lifetime (approximately equal to the atomic excited state lifetime), the method guarantees that the atom is in the $F=3$ hyperfine ground state immediately after the detection of the second photon.  In principle information on the $m$-state could be gained through initial optical pumping and increased Zeeman separation of levels to distinguish the transitions in frequency.  However, this would also require clearly choosing a detection method between Faraday rotation (no change in $m$) and spontaneous emission collection ($\Delta m=\pm 1$).  In the regime where we have demonstrated detection, the combination of the two processes leaves the atom in a mixed state of $m$-sublevels after the photon detections. This technique is a classical determination of atom presence rather than a projection onto a pure quantum state.

One significant advantage of this scheme is that it is sensitive enough to work without strong coupling of the cavity and atoms, meaning that the detection volume can be kept large.  As presented in Eq. \ref{xperp2}, the signal rate for both spontaneous emission collection and Faraday rotation (for small rotations) depends on the square of the single-atom cooperativity, which is proportional to the cavity finesse.  For a given mode volume, increasing the finesse by some factor (through higher reflection coefficients on the mirrors) will in general increase the detection efficiency.  However, the cavity linewidth decreases by the same factor, which may reduce the rate of escape of light from the cavity.
The signal-to-noise ratio of Faraday rotation to birefringence is independent of the finesse, since both arise from a rotation of the polarization on each round trip of the light.

We have demonstrated how autocorrelation measurements of the emitted light show the underlying dynamics of the atom as it traverses the Gaussian mode of the cavity and encounters the standing wave.
This method of atomic detection, using a conditional measurement, is well-suited for some quantum control protocols. The implementation should facilitate the manipulation of the electromagnetic field produced in the atom-mode interaction of cavity QED systems.

\section{Acknowledgments}
This work was supported by the National Science Foundation. We are grateful to PicoQuant Photonics
for their loan of the PicoHarp 300 time-correlated single photon counting module.
We thank interactions with  M. L. Terraciano, J. Jing, R. Olson Knell and A. Fern{\'a}ndez at the early stages of this work.

\end{document}